\documentstyle[amssymb,12pt]{article}

\input tcilatex

\begin{document}

\centerline{\bf BUBBLE EXPANSION  AT THE END OF INFLATION}

\bigskip \centerline{ S.G.Rubin$^{1,2}$}

$^1$Moscow Engineering Physics Institute,

$^2$Center for Cosmoparticle Physics - ''Cosmion''

\centerline{ABSTRACT}

A dynamic of a false vacuum decay is investigated at the end of inflation.
It is shown that a creation of a new phase is mainly stimulated by
temperature fluctuations. Colliding walls of O(3) - symmetrical bubbles
interact weakly with surrounding that leads to large field fluctuations.

\centerline{INTRODUCTION}

An investigation of a temperature aspect of the Universe evolution and its
influence a modern structure and distribution of a matter has a long
history. Nevertheless the question on the temperature of the Universe at
early stages of evolution is still opened. The evolution of the Universe in
this case could be rather different from that with zero temperature \cite
{Barr}. Both cases suppose a presence of scalar field that cause an
inflation - a necessary tool of modern scenarios of an evolution of early
Universe. There are several ways to choose a Lagrangian of the scalar
(inflaton) field. If it possesses more than one minimum the first order
phase transitions could take place. This transition proceeds by a nucleation
of an O(3) - symmetrical bubbles with true vacuum inside it in the heated
Universe and O(4) - symmetrical bubbles in the case of zero temperature.
Both sorts of the bubbles can produce particles and large scale fluctuations
of the inflaton field during collisions of expanded bubbles.

The difference in the original symmetries leads to a meaning difference in a
destiny of the large fluctuations. The O(4) - symmetrical bubbles produce a
discal matter distribution in an area of their collision . As it was shown
in \cite{Wu} black holes are not produced in this case if one neglects the
radiation. The matter distribution after the temperature decay of the false
vacuum is rather different as well. As it was shown in \cite{KRSK}, there is
a real possibility to form black holes during the bubble collision, a number
of which is large enough to leave the observable effects. The aim of this
paper is to investigate conditions of a dominance of the temperature decay
at the end of inflation and the interaction of expanded bubbles with
surroundings.

\centerline{ DOMINANCE OF TEMPERATURE DECAYS}

It is well known , see for example \cite{Dymn}, that a quantum evaporation
of particles in a presence of a horizon could be responsible for the
Universe heating at the stage of inflation. The temperature of the Universe
is connected with horizon size, $T_{H}=H/2\pi $ for $H=Const$, similar to
Hawking formula for the evaporation of black holes \cite{Hawk}. If the size
of the horizon increases with time, it leads to a cooling of the Universe
and indicates an absence of an equilibrium in the system. As every system,
the Universe possesses some inertia and hence its average temperature should
be grater than instantaneous temperature $T_{H}$:

\begin{equation}  \label{Thor}
T\geq T_H=H/2\pi
\end{equation}

Consider the false vacuum decay due to temperature fluctuations at the end
of inflation and conditions of its prevail over the tunneling decay. The
temperature fluctuations create O(3) - symmetrical bubbles with true vacuum
of a scalar field inside it whereas O(4) - symmetrical bubbles are formed by
tunneling. The probability of the O(3) - symmetrical bubble formation is 
\cite{Linde}:

\begin{equation}  \label{Ptemp}
P_{temp}\propto \exp (-S_3/T),
\end{equation}
where $T$ is the temperature of the phase transition, $S_3$ is an Action of
O(3) - symmetrical solution to the classical equation of motion for the
scalar field $\varphi $ with a Lagrangian:

\begin{equation}
\begin{array}{c}
L=\frac{1}{2}(\partial _{\mu }\varphi )^{2}-V(\varphi ), \\ 
V(\varphi )=\frac{\lambda }{8}(\varphi ^{2}-\varphi _{0}^{2})^{2}+\epsilon
\varphi _{0}^{3}(\varphi +\varphi _{0}).
\end{array}
\label{Lagr}
\end{equation}
The parameters of the Lagrangian depend on temperature, but one can neglect
it because the first order phase transition supposed to be quick enough.
This form of the potential is quite general and used by many authors (see
for example, \cite{Watkins}) to investigate the false vacuum decay at zero
temperature. Another wide-spread form of the potential contains cubic power
of the inflaton field rather than linear term. One can easily check that
both forms are equivalent in the thin wall approximation by an appropriate
shift of field $\varphi .$

There is no reason for the linear (or cubic) term to disappear with the
temperature variation. An analytic results can be obtained in the framework
of 'thin wall approximation' \cite{Cole} - $\epsilon $$/\lambda <<1$ and we
will use it in the following. The false vacuum at $\varphi =\varphi _{0}$
decays by the formation of bubbles with the true vacuum state $\varphi
=-\varphi _{0}$ inside it. The situation like this is realized for example
in the theory of thermal inflation \cite{Barr}. \qquad

Compare the probability of the temperature phase transition (\ref{Ptemp})
and tunnel one that has the form \cite{Cole} 
\begin{equation}
P_{tun}\propto \exp (-S_{4}).  \label{Ptun}
\end{equation}
Here $S_{4}$ is the Action for the O(4) - symmetrical solution. The
condition of the dominance of the temperature phase transition is evidently $%
S_{3}/T<S_{4}$. The values $S_{3}$ and $S_{4}$ can be calculated in the thin
wall approximation \cite{Cole}, \cite{Linde} and one can easily come to the
inequality 
\begin{equation}
T\geq \frac{32}{27\pi }\frac{\epsilon }{\lambda }.  \label{limit1}
\end{equation}
Here and in the following the system of units $\hbar =c=k_{B}=m=1$ is used.
The mass of the scalar particles of the field $\varphi $ is $m=\sqrt{\lambda 
}\varphi _{0}$.

The inequality (\ref{limit1}) can be expressed now in terms of the
Lagrangian parameters. Equating the experimental value of the horizon size
at the end of inflation, $H_{exp}\approx 4\cdot 10^{-6}m_{pl}$ and its
theoretical value $H_{theor}=(8\pi \rho _{V}/3m_{pl}^{2})^{1/2}$, where $%
\rho _{V}=2\epsilon m^{4}/\lambda ^{2}$ is the energy density of the false
vacuum state, we obtain the normalization: $m_{pl}/m\approx 10^{3}(\epsilon
/\lambda ^{2})^{1/4}$. The second limit for the temperature of the Universe
at the moment of the transition can be obtained after simple substitution: 
\begin{equation}
T\geq 10^{-3}\frac{\epsilon ^{1/4}}{\lambda ^{1/2}}  \label{limit2}
\end{equation}
($T\sim 10^{16}\epsilon /\lambda (GeV)$ in more conventional units).

The third limit is the consequences of causality - the initial size of the
bubble $R_{0}=2\lambda /3\epsilon $ should be less than horizon size $H^{-1}$
- and is not connected with the temperature: 
\begin{equation}
\frac{\epsilon ^{3/4}}{\lambda ^{1/2}}\geq 4\cdot 10^{-3}  \label{limit3}
\end{equation}
Inequalities (\ref{limit1}), (\ref{limit2}) and (\ref{limit3}) do not
contradict each other in a wide range of the parameter values and one can
conclude that if the inflation is terminated by the first order phase
transition it would be the result of the temperature fluctuations rather
than the tunneling.

The bubble of true vacuum is separated from the false vacuum surrounding by
the walls at rest. After nucleation the wall starts to accelerate outwards
absorbing false vacuum region and converting the difference of false and
true vacuum energy density into a kinetic energy of the wall. This process
continues up to the collision with a spherical wall of another bubble. The
kinetic energy of the walls is transformed into matter radiation and large
scale fluctuations of the scalar field. The last take place if the kinetic
energy of the walls is large enough but, being produced, leads to meaning
consequences like a multiple production of black holes \cite{KRSK}. This
conclusion does not depend on the form of the Lagrangian provided the two
minimums exist and thin wall approximation is correct.

The rest of the paper is devoted to the problem of the wall propagation in
the case of the phase transition at the end of the inflation. It is shown
that the walls interact weakly with the particles and their kinetic energy
increase as if there is no friction.

\centerline{WALL MOVEMENT TROUGH THERMAL BACKGROUND}

Two factors effect on the kinetic energy of the colliding walls: a distance
between the bubble centers and an interaction of the walls with thermal
plasma. As it was mentioned above, the horizon heats the Universe by the
evaporation of all possible sorts of particles. It is the interaction with
these particles that leads to a friction and made the walls move not very
quickly. For example the walls move with the velocity $\symbol{126}~0.5$ in
the case of late phase transitions \cite{Turok}.

Let us investigate separately the scattering of the inflaton particles and
other 'light' particles with the masses $m<<T$ \ by the wall at rest. It is
well known that there are two regimes in dependence whether a length free
path of the particles is much more or much less than the wall width $L$ \cite
{Turok}. The wall width can be estimated easily: $L\approx 1/m=1$ in chosen
units. The main uncertainty is contained in the length free path that
depends on cross sections. To estimate them suppose the interaction in the
form: 
\begin{equation}
L_{int}=g_{1}\varphi \psi ^{+}\psi +g_{2}\varphi ^{2}\phi ^{2}+g_{3}\phi
\psi ^{+}\psi ,  \label{Li}
\end{equation}
where $\psi $ is an operator of spinor particles, $\phi $ is an operator for
scalar particles, other than that is responsible for the inflation. The
interaction of the inflaton (scalar) particles can be obtained from Eq. (\ref
{Lagr}) and equals $\lambda \varphi ^{4}$. No symmetries are assumed to be
broken yet and all the constants are supposed of the same order $g_{1}\sim
g_{2}\sim g_{3}\sim \lambda <<1$. The temperature $T\sim \epsilon /\lambda $
(see(\ref{limit1})) is small comparing the mass of inflaton quanta and large
comparing the masses of other particles. The interaction of light particles
leads to the largest cross section that can be estimated as $\sigma $$\sim
g_{3}^{4}/T^{2}$. Here the expression $E_{p}\approx 3T$ for the energy of
relativistic particles at the temperature $T$ is presumed.

The length free path equals to $l=1/\sigma n$, where $n=\kappa \sigma
_{s}/3\cdot T^{3}$ is a number of the light particles at the temperature $T$ 
\cite{Khlo}. The constant $\sigma _{s}=\pi ^{2}/15$ in our units and a total
number of sorts of the light particles is $\kappa \sim 50.$ Thus one can
easily check that the length free path $l\sim 1/(10g_{3}^{4}T)$ is much more
than the wall width $L\approx 1$ (remind that we are working in the limits $%
g_{3},\epsilon /\lambda <<1$ ). Therefore we can consider the interaction of
the wall with the particles supposing them to be free, on the contrary to
the electroweak transition \cite{Turok}.Another essential difference
consists of the equality of the particle masses at both sides of the wall
because the inflaton field is not responsible for the mass generation.

To estimate the strength of the interaction at the end of inflation we
neglect further the asymmetry of the potential $V(\varphi )$, that is
proportional to the small parameter $\epsilon $$/\lambda $, and a curvature
of the wall.

To describe the scattering of inflaton particles one can consider them as
small fluctuations around classical solution $\varphi =\varphi _{cl}+\varphi
^{\prime }$, where $\varphi _{cl}=\varphi _{0}th(z/2)$ is the flat wall
solution to the classical equation of motion 
\begin{equation}
\Box \varphi +\frac{\partial V}{\partial \varphi }=0  \label{Cleq}
\end{equation}
After a linearization Eq.(\ref{Cleq}) has the form \cite{Raj} 
\begin{equation}
\left( \frac{d^{2}}{dz^{2}}+4\omega ^{2}-4+\frac{6}{ch^{2}z}\right) \varphi
^{\prime }=0.  \label{Lineq}
\end{equation}
The one-particle scattering by the wall with the particle energy $\omega $
and momentum ${\bf k}$ should satisfy the boundary conditions $\varphi
^{\prime }(t\rightarrow -\infty ,z)=\exp (-i\omega t+ikz)$ and $\varphi
^{\prime }(t\rightarrow \infty ,z)=A\exp (-i\omega t+ikz)+B\exp (-i\omega
t-ikz)$. It can be shown that the solution to Eq.(\ref{Lineq}) is
reflectionless \cite{Land}, i.e. $ReA=1,B=0$ and hence the momentum transfer
equals zero. Scalar particles do not scatter by a flat wall of the same
scalar field and do not slow down the wall.

The only what remains is to estimate the pressure of the other particles:

\begin{equation}
p_{p}=qnW.  \label{pp1}
\end{equation}
Here $q$ is the momentum transfer, $n=10T^{3}$ is the particle density and $%
W $ is the probability of the scattering of the incoming particle. The wall
at rest can be considered as an external field and a conservation of an
energy $E_{k}$ and parallel projection of the particle momentum ${\bf k}$
determines the momentum transfer: ${\bf q}=(0,0,-2k_{z})$. The probability $%
W $ can be expressed in the form $W=\left| M_{k,k+q}\right|
^{2}/(8k_{z}E_{k})$ in the case of planar wall perpendicular to $z$ axis.
The matrix element is obtained using the first term of expression (\ref{Li})
where only classical part $\varphi _{cl}$ of the field $\varphi $ are taken
into account :

\begin{equation}  \label{Matr}
M_{k,k+q}=g_1 \varphi _0\int_{-\infty}^\infty dze^{iqz}th(z/2)
\end{equation}

The integral is easily estimated if one substitutes a linear function for $%
th(z/2)$\ in the main region $|z|<2$: 
\[
M_{k,k+q}=i2g_{1}\varphi _{0}\frac{\cos 2q-\sin 2q/q}{q}\leq i2g_{1}\varphi
_{0} 
\]
The last inequality takes into account that the $q$-dependent ratio is less
than approximately one. So expression (\ref{pp1}) becomes 
\[
p_{p}\leq n\frac{g_{1}^{2}\varphi _{0}^{2}}{E_{k}}. 
\]
The energy of the incoming particle in the rest frame of the wall is $%
E_{k}\sim \gamma T$ and the upper limit of the pressure can be written as 
\begin{equation}
p_{p}\leq 10\frac{g_{1}^{2}}{\gamma \lambda }\left( \frac{\epsilon }{\lambda 
}\right) ^{2}  \label{pp2}
\end{equation}

The incoming particles cause the pressure $p_{p}$ that decelerates the wall.
On the other side the wall is accelerated by a pressure difference $%
p_{\varphi }=\rho _{V}=2\epsilon /\lambda ^{2}$ in the true and false
vacuums that are separated by the wall. Its ratio is given by

\begin{equation}
\frac{p_{p}}{p_{\varphi }}\leq \frac{g_{1}^{2}}{\gamma }\frac{\epsilon }{%
\lambda }  \label{ratio}
\end{equation}
This ratio is always much less then unity because both $g_{1}$ and $\epsilon 
$$/\lambda $ are supposed to be small values. Note that expression (\ref
{ratio}) is the upper limit for the pressure. As it follows from the form of
the matrix element $M_{k,k+q}$, the pressure tends to zero not only at small
momentum transfer but at large one as well. The same estimation of upper
limit of the pressure could be done for the scattering of light scalar
particles. The conclusion on the absence of the friction is in agreements
with that for electroweak transition \cite{Turok} in the limit of the
particle mass equality on the both sides of the wall. In our case the masses
are equal because the absolute value of the fields in the two minimums are
approximately equal. If the Lagrangian has the local minimum at $\varphi =0$
the friction would appear. The similar situation was considered in \cite
{Dine}.

Thus the friction of the wall that moves through the heated medium at the
end of inflation is small enough and the bubble walls collide having large
kinetic energy. It could be in its turn the reason of the large density
fluctuation of the inflaton field.

\centerline{CONCLUSION}

In this paper the problem of the first order phase transitions due to
temperature fluctuations at the end of inflation is investigated. It was
shown that the temperature vacuum decay prevail the tunneling one in wide
range of the parameters. The walls of the expanding true vacuum bubbles are
almost transparent for the thermal plasma unlike the case of electroweak
first order phase transition. The walls collide being highly relativistic.

The nucleated O(3) - symmetrical bubbles of the true phase are expanded with
acceleration up to the moment of the collision of their walls. The last
leads to the large scale fluctuations of the inflaton field together with
ordinary radiation that reheated the Universe. The large scale fluctuations
results to a number of observable effects and as it was shown in \cite{KRSK}
are able to collapse thus forming black holes. This mechanism of the
production of the primordial black holes is rather effective to influence
the Universe evolution.

\centerline{ACKNOWLEGMENTS}

The author is grateful to M. Yu. Khlopov for the fruitful discussions and to
R. V. Konoplich and A. S. Sakharov for their interest to the work.

This work was undertaken in the scientific and educational center
''Cosmion'' in the framework of Section ''Cosmomicrophysica'', of State
program ''Astronomy. Fundamental Space Research'' and International project
''Astrodamus''.

\end{document}